\documentclass[10pt,twocolumn]{revtex4}
\usepackage{graphicx}
\usepackage{amsmath}
\usepackage{amssymb}
\usepackage{}
\begin{document}

\preprint{APS/123-QED}


\title{Helicity reversion in high harmonic generation driven by bichromatic counterrotating circularly polarized laser fields}

\author{Xiaofan Zhang$^{1}$}
\author{Liang Li$^{1}$}

\author{Xiaosong Zhu$^{1,}$\footnote{Corresponding author: zhuxiaosong@hust.edu.cn}}

\author{Xi Liu$^{1}$}
\author{Qingbin Zhang$^{1}$}

\author{Pengfei Lan$^{1}$}

\author{Peixiang Lu$^{1,2}$}

\affiliation{$^1$School of Physics, Huazhong University of Science
Technology, Wuhan 430074, China\\
$^2$Laboratory of Optical Information Technology, Wuhan Institute of
Technology, Wuhan 430205, China}

\date{\today}

\begin{abstract}
We investigate the polarization properties of high harmonics generated with the bichromatic counterrotating circularly polarized (BCCP) laser fields by numerically solving time-dependent Schr\"{o}dinger equation (TDSE). It is found that, the helicity of the elliptically polarized harmonic emission is reversed at particular harmonic orders. Based on the time-frequency analysis and the classical three-step model, the correspondence between the positions of helicity reversions and the classical trajectories of continuum electrons is established. It is shown that, the electrons ionized at one lobe of laser field can be divided into different groups based on the different lobes they recombine at, and the harmonics generated by adjacent groups have opposite helicities. Our study performs a detailed analysis of high harmonics in terms of electron trajectories and depicts a clear and intuitive physical picture of the HHG process in BCCP laser field.
\end{abstract}                         
\maketitle

\section{Introduction}

High-harmonic generation (HHG) occurs in the interaction of intense femtosecond laser field with atoms or molecules \cite{PBCorkum,FKrausz}. HHG has been intensively studied in experimental and theoretical investigations \cite{Zhu2,Xin,Zhu3}, because it promises both applications for generating attosecond pulses \cite{Sansone,LM,Cavalieri} and for probing the molecular structure \cite{Itatani,Haessler,Kraus,Zhai} and ultrafast dynamics \cite{Smirnova,Baker,HJW,Travnikova,Boeglin}. The HHG process can be nicely explained by the three-step model \cite{Corkum,Schafer}. First, under the influence of the laser field, an electron escapes from the atom or molecule and reaches the continuum. Then it is accelerated in the laser field. Finally the accelerated electron is driven back to the parent ion and a high energy photon is emitted. According to this picture, only laser fields, which are able to drive the electrons back to the ion, can be used to efficiently generate high harmonics.

Very recently, the bicircualr counter-rotating circularly polarized (BCCP) driving laser field has been a hot spot due to its unique features and applications. A BCCP laser field consists of two coplanar counter-rotating circularly polarized fields with the fundamental field frequency $\omega$ and its second harmonic $2\omega$ \cite{Long,Eichmann,DB}. The time-dependent electric field performs a Lissajous figure on its polarization plane with a three-fold symmetry, i.e. it consists of three segments or lobes per cycle. Therefore, the laser field peaks three times one cycle of the fundamental, producing three ionization bursts. The electrons ionized to the continuum near the laser peaks can successfully recombine with the parent ion, radiating an attosecond pulse train with three pulses one cycle of the fundamental \cite{LM,DB2,DB3,DB4}.

This attractive laser field has been demonstrated to be an effective tool for fully controlling the high harmonic polarization from linear through elliptical to circular polarization \cite{Fleischer}. Importantly, the conversion efficiency for harmonics with arbitrary polarization is comparable to that of the standard HHG process, which yields linearly polarized high harmonics driven by a linearly polarized laser pulse \cite{LM,DB,Fleischer}. When the BCCP laser pulse is acting on an isotropic medium, in the frequency domain, a comb of high efficiency harmonic with frequency (3n$\pm$1)$\omega$ is radiated \cite{LM,Kfir,Reich,AF}. According to the selection rules \cite{Fleischer,Alon,Pisanty,DB5,Mauger}, the harmonics with frequency (3n+1)$\omega$ have the same helicity as the fundamental while the (3n-1)$\omega$ harmonics have the same helicity as the second harmonic pulse (3n$\omega$ is parity forbidden). Selecting harmonics with one kind of helicity would generate attosecond pulses with circular polarization. Some efforts have been made for the selection, e.g., optimizing the phase-matching conditions \cite{Kfir} or using crossed beams of counter-rotating circularly polarized driving lasers \cite{Hickstein}. This circularly polarized light source is very useful for ultrafast spectroscopy, especially for the ultrafast circular dichroism studies \cite{AF,Hergenhahn,NB,Powis}. It was also shown that the bicircular HHG driven by this BCCP laser pulse can be a new method to probe dynamical symmetries of atoms and molecules and their evolution in time \cite{Baykusheva,Reich2}. These attractive features of HHG in BCCP laser fields are related to the symmetry of the laser field and the corresponding three-step dynamics of the continuum electrons. To better understand the features of HHG in the BCCP laser field, the correspondence between the harmonic polarization and the electron trajectories needs to be established.

In our study, the polarization properties of high harmonics generated in 2D model Ne atoms by the BCCP laser fields are investigated. We focus on the helicity reversions in the harmonic spectrum, where the rotation direction of the harmonic electric field reverses at specific orders. To explain this phenomenon, we depict the three-step picture of HHG in BCCP fields and reveal the correspondence between the time-frequency distribution of harmonic degree of circular polarization and the classical trajectories. Our study helps people better understand the HHG process and depicts a clear picture for the HHG process in the BCCP laser field.

\section{Theoretical model}

The high harmonic spectrum generated from the interaction between the BCCP laser field and Ne atoms is calculated by numerically solve the two-dimensional time-dependent Schr\"{o}dinger equation (TDSE) in length-gauge \cite{M.,Rui}.
The 2D model potential of Ne atom is given by \cite{Barth,Zhu1} (atomic units(a.u.) are used throughout this paper):
\begin{eqnarray}
V(\vec{r})= - \frac{Z(\vec{r})}{\sqrt{\vec{r}^{2}+a}},
\end{eqnarray}
where $Z(\vec{r})=1+9\exp(-\vec{r}^{2})$ and a=2.88172 to obtain the ionization potential of Ne atom $\textit{I}_{p}$=0.793 a.u. for 2$\textit{p}$ orbitals. $\vec{r}$$\equiv(x,y)$ denotes the electron position in the two-dimensional x-y plane.

The bicircular electric field is obtained by combining left-circularly polarized and right-circularly polarized laser fields. The laser vector in the $x-y$ polarized plane is defined by \cite{LM}
\begin{eqnarray}
E(t)=\nonumber E_{0}f(t)(\cos[\omega t]+\cos[\gamma\omega t])\hat{x} \\ +E_{0}f(t)(\sin[\omega t]-\sin[\gamma\omega t])\hat{y},
\end{eqnarray}
where $\gamma$ denotes the integer multiple of the fundamental frequency. The laser field has $\gamma+1$ laser lobes and reflects $\gamma+1$ fold symmetry. The $\omega$ field with helicity +1 rotates counterclockwise while the second field $\gamma\omega$ with helicity -1 rotates clockwise. $f(t)$ is the trapezoidal envelop with 2 cycle rising and 2 cycle falling edges and 5 cycle plateau (in units of fundamental). We set the same laser intensity between the $\omega$ field and the $\gamma\omega$ field. $E_{0}$ is the amplitude of the laser field.

We use the split-operator method to solve Eq.(1). To avoid spurious reflections from the spatial boundaries, the electron wave function $\Psi(\vec{r},t)$ is multiplied by a ``mask function" $g$ with the following form \cite{Yuan}
\begin{eqnarray}
g=g(x)g(y)
\end{eqnarray}
at each time step, in which
\begin{eqnarray}
g(x)=\{^{1\qquad\qquad\qquad\ \ \ |x|<R_{x}-L_{x}} _{\sin^{1/8}(\frac{R_{x}-|x|}{2\pi L_{x}})\qquad|x|\geq R_{x}-L_{x}}
\end{eqnarray}
and
\begin{eqnarray}
g(y)=\{^{1\qquad\qquad\qquad\ \ \ |y|<R_{y}-L_{y}} _{\sin^{1/8}(\frac{R_{y}-|y|}{2\pi L_{y}})\qquad|y|\geq R_{y}-L_{y}}.
\end{eqnarray}
For all results reported here, we set the width of the ``absorbing" area as $L_{x}$=60 a.u. and $L_{y}$=60 a.u.. In our calculation, we set the range of the ``mask function'' from -240 a.u. to 240 a.u. in the Cartesian coordinates, i.e., $R_{x}$=$R_{y}$=240 a.u.. The space step is 0.1 a.u.. The time step is $\Delta t=0.05$ a.u..

The time-dependent dipole acceleration can be obtained by means of the Ehrenfest theorem
\begin{eqnarray}
A_{q}(t)=-<\Psi(t)|[H(t),[H(t),q]]|\Psi(t)>, q=x,y .
\end{eqnarray}

Then, the harmonic spectrum is obtained by Fourier transforming the dipole acceleration
\begin{eqnarray}
a_{q}(\omega)=\int A_{q}(t)\exp(-iq\omega t)dt.
\end{eqnarray}
The intensity of left- and right- polarized harmonic components can be obtained by
\begin{eqnarray}
D_{\pm}=|a_{\pm}|^{2},
\end{eqnarray}
where $a_{\pm}=\frac{1}{\sqrt{2}}(a_{x}\pm ia_{y})$. The polarization state of high harmonics can be characterized by the degree of circular polarization (DCP) \cite{DB5,Antoine,Hecht,Born,DB6}
\begin{eqnarray}
 \zeta=\frac{D_{+}-D_{-}}{D_{+}+D_{-}},
\end{eqnarray}
which can be obtained from the intensity measurements.
The ellipticity $\epsilon$ has a one-to-one correspondence with the DCP as $\zeta=\frac{2\epsilon}{1+\epsilon^2}$ \cite{DB}. The two parameters both vary in the interval from -1 to 1 and have the same sign.

The rotation direction of the electric field can be quantitatively described by the helicity, which is defined as:
\begin{eqnarray}
h=sgn(\epsilon)=sgn(\zeta).
\end{eqnarray}
The helicity $h$ takes the values -1 and +1, indicating the two opposite rotation directions.

\section{Results and discussions}

\begin{figure*}
\centerline{
\includegraphics[width=15cm]{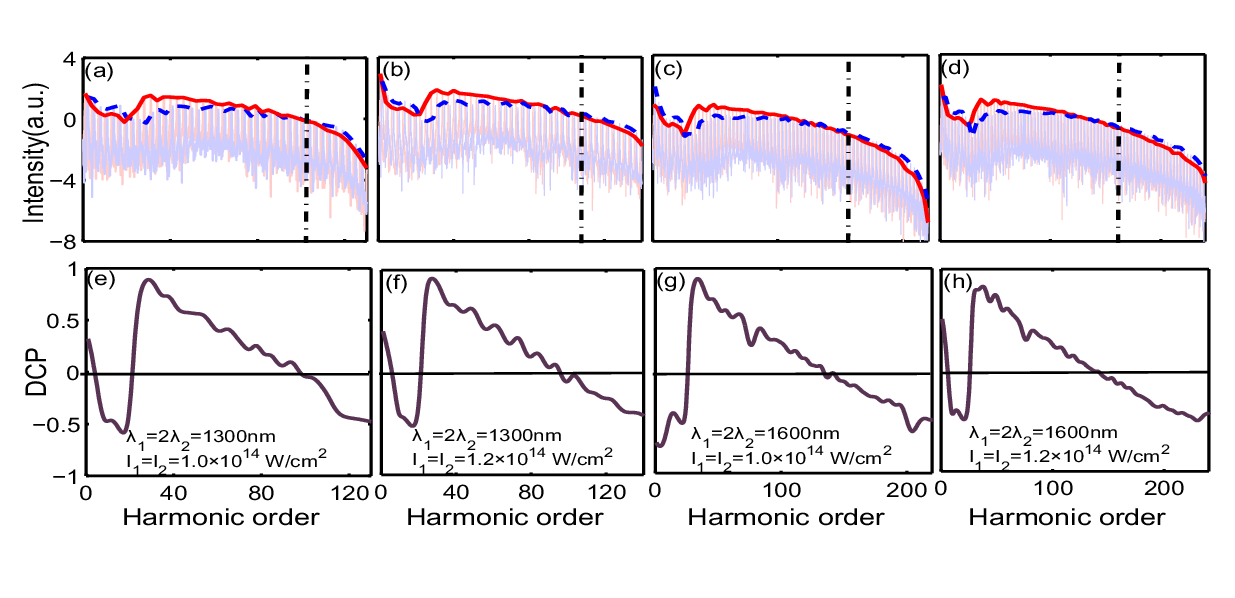}}
\caption{(a)-(d)Harmonic intensity (e)-(f) DCP as a function of the harmonic order for HHG from the 2$\textit{p}$ orbital of Ne atoms driven by bicircular $\omega-2\omega$ laser field with different laser parameters. The fundamental laser wavelengths and the laser intensities are: (a) and (e) $\lambda_{1} =$ 1300 nm and $\textit{I}_{1} = \textit{I}_{2}$ = 1.0 $\times 10^{14}$ W/cm$^{2}$, (b) and (f) $\lambda_{1}$ = 1300 nm and $\textit{I}_{1}= \textit{I}_{2}$ = 1.2 $\times 10^{14}$  W/cm$^{2}$, (c) and (g)  $\lambda_{1}$ = 1600 nm and $\textit{I}_{1}= \textit{I}_{2}$ = 1.0 $\times 10^{14}$ W/cm$^{2}$, (d) and (h) $\lambda_{1}$ = 1600 nm and $\textit{I}_{1}= \textit{I}_{2}$ = 1.2 $\times 10^{14}$ W/cm$^{2}$. The spectra for harmonics co-rotating and counter-rotating with the fundamental field are shown as the light red and blue lines respectively. The envelopes of these harmonic spectra are presented as the dark solid red and blue dashed lines correspondingly.}
\end{figure*}


Firstly, we present the harmonic intensity and the DCP as a function of the harmonic order with different laser parameters in Fig.1. We choose the example of Ne atoms with the 2$\textit{p}$ ground state. In this case, the two components of the bicircular field with $\gamma$=2 have equal intensities. According to the selection rules, the harmonics 3n$\omega$ are absent, and the harmonics (3n+1)$\omega$ and (3n-1)$\omega$ have opposite helicities. The harmonic spectra for left-circularly polarized harmonics (3n+1)$\omega$ and right-circularly polarized harmonics (3n-1)$\omega$ are presented with light red lines and blue lines respectively. The envelopes of these harmonic spectra are presented with the dark red solid lines and blue dashed lines correspondingly. For equal intensities of the two laser field components, the position of the harmonic cutoff is obtained by \cite{DB}
\begin{eqnarray}
n\omega= \frac{3.17U_{p}}{\sqrt{2}}+1.2|I_{p}|.
\end{eqnarray}
The ponderomotive energy of the entire field is defined by $\textit{U}_{p}=\frac{2E_{0}^2}{4\omega^2}+\frac{2E_{0}^2}{4(2\omega)^2}=\textit{U}_{p1}+\textit{U}_{p2}$. The cutoff obtained by Eq.(11) is almost coincident with that from the simulated results.

The harmonic spectrum driven by the bicircular laser field with the fundamental laser wavelength $\lambda_{1}=1300$ nm and laser intensity $\textit{I}_{1}=1.0\times 10^{14}$ W/cm$^{2}$ is shown in Fig. 1(a). From this figure, one can see that, the intensities of the harmonics (3n-1)$\omega$, having the helicity $h=-1$, are different from that of the harmonics (3n+1)$\omega$, for which $h=+1$. There are two helicity reversions where the red-solid and blue-dashed lines intersect. One is at about the ionization threshold (22nd-order harmonic), the other is at about 107th-order harmonic (labelled by the dash-dot line). Below the ionization threshold, the intensities of (3n+1)th-order harmonics are higher than those of (3n-1)th-order. Between the ionization threshold 22nd-order and the 107th-order harmonic, the intensities of (3n-1)th-order harmonics turn to be higher than those of the (3n+1)th-order. While between 107th-order and the cutoff 120th-order obtained by the Eq.(11), an opposite result to that between the ionization threshold 22nd-order and the 107th-order harmonic order appears. That is, the intensities of the harmonics (3n+1)th-order co-rotating with the $\omega$ field are higher than those of the (3n-1)th-order counter-rotating with the $\omega$ field. In order to clearly show the helicity reversion phenomenon in the harmonic spectra, we display the DCP in Fig.1(e). According to the helicity definition, the helicity reversion appears at the position where the DCP changes its sign. From Fig.1(e), The DCP crosses zero at ionization threshold 22nd-order and 107th-order harmonic, which is coincident with the discussion above.

To check whether the phenomenon is general for the bicircular field interaction with Ne atoms and uncover the reason for the reversion, we apply different laser parameters for fundamental laser wavelength $\lambda_{1}=1300$ nm , 1600 nm and laser intensity $\textit{I}_{1}=1.0\times 10^{14}$ W/cm$^2$, $1.2\times 10^{14}$ W/cm$^2$. The harmonic spectra are presented in Figs.1(b-d) and the DCP are plotted in Figs.1(f-h). It is shown that the intensities of (3n+1)th-order harmonics are higher than those of the (3n-1)th-order harmonics below the ionization threshold, and correspondingly, the values of the DCP are negative. There is a helicity reversion at the ionization threshold for different laser parameters. We have also calculated the harmonic spectra for other model atoms with different two-dimensional soft-core potentials and different ionization potentials. It is shown that the reversion at the threshold is not a general phenomenon and is dependent on the target chosen. Except the helicity reversion at the ionization threshold, the helicity reversion above the threshold is found in all these spectra. The reversion position is labelled by the dash-dot line. In our study, we mainly investigate the helicity reversions in the above-threshold region.

\begin{figure}
\centerline{
\includegraphics[width=8.5cm]{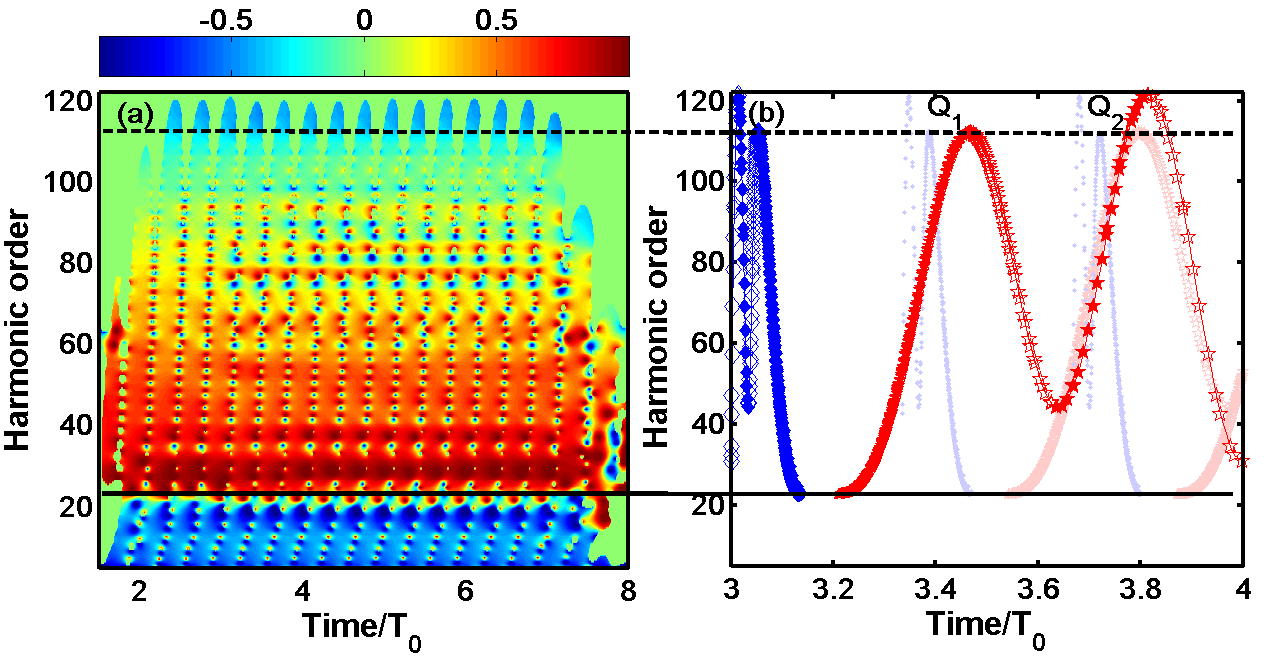}}
\caption{(a) The time-frequency distribution of the harmonic DCP. (b) The classical paths in the BCCP laser field. The blue diamonds and red pentagrams present the ionization time and the recombination time. The solid markers denote the short trajectories and the hollow markers denote the long trajectories. The laser parameters are the same as in Fig.1(a). $T_{0}$ is the optical cycle of the $\omega$ laser field. The horizontal solid line marks the $\textit{I}_{p}$ and the dashed line marks the position of the helicity reversion.}
\end{figure}

In order to explain the phenomenon of the helicity reversion and elucidate the physics behind it, we present the time-frequency distribution of the harmonic DCP and the classical ionization and recombination paths in Fig.2. The laser parameters are the same as in Fig.1(a). Here we analyze the time-frequency properties of the harmonic DCP with the Gabor transform, which has been proved to be a very powerful tool to analyze the emission times of HHG and to clearly discriminate the electron trajectories. Considering the dipole acceleration $A_{q}(t)$ of Eq.(6), the Gabor transform is defined as \cite{CC}

\begin{eqnarray}
GT_{q}[\Omega,t_0]=\frac{1}{2\pi}\int dtA_{q}(t)e^{-i\Omega t}e^{-(t-t_{0})^2/2\sigma^2},
\end{eqnarray}
where $\Omega$ is the frequency of the high harmonics. In our study, we use $\sigma=1/3\omega$. The DCP distribution is obtained by $\zeta=\frac{|GT_{+}|^2-|GT_{-}|^2}{|GT_{+}|^2+|GT_{-}|^2}$, in which $GT_{\pm}=\sqrt{\frac{1}{2}}(GT_{x}\pm iGT_{y})$. In Fig.2(a), we present the DCP distribution of harmonics between 1.5 optical cycle and 8 optical cycle. Due to the symmetry of the BCCP laser pulse, three radiation bursts per fundamental laser pulse cycle can be found in this figure, as predicted in previous works \cite{LM,DB2,DB3,DB4}. There are two helicity reversions (labelled by the sold line and the dashed line respectively) and three distinct regions can be identified: (\uppercase\expandafter{\romannumeral1}) Below threshold 22nd-order (labelled by the solid line), the DCP is mostly negative. The harmonic emissions counter-rotating with the $\omega$ laser field are radiated. (\uppercase\expandafter{\romannumeral2}) The middle region between the threshold (22nd-order) and the second reversion (about 107th-order, labelled by the dashed line), the DCP is high and positive. The harmonic emissions are emitted with the left-circular polarization. (\uppercase\expandafter{\romannumeral3}) The upper region between the second reversion (about 107th-order) and the cutoff (120th-order), the DCP is almost negative opposed to the second region. In the second region, the highest DCP is at about the 30th-order harmonic, corresponding to the position of the maximum intensity difference between the left-circularly and right-circularly polarized harmonic components in Fig.1(a). Then the pulse DCP decreases gradually along with the increase of the harmonic order. At about 107th-order denoted by the dashed line, the DCP transits from positive value to negative value. These are directly reflected in Fig.1(e).

\begin{figure}
\centerline{
\includegraphics[width=8.5cm]{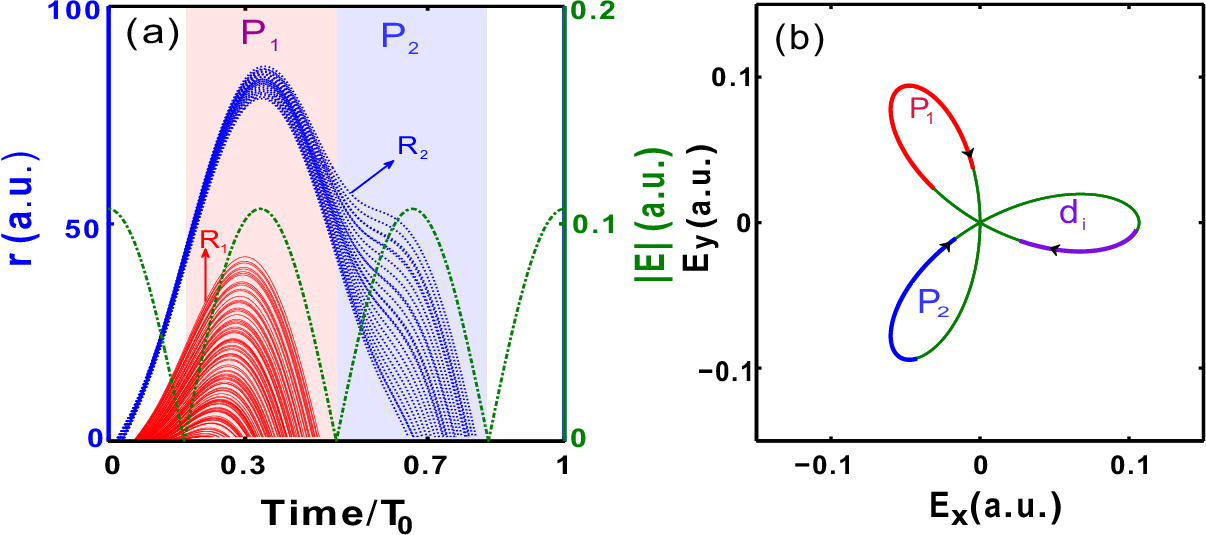}}
\caption{(a) The amplitude of the electronic field (green dash-dot line, right vertical axis) and the trajectories of the electrons ionized at the first lobe and recombining with the parent ion at the following lobes. The second laser lobe $P_{1}$ and the third laser lobe $P_{2}$ are superimposed by the light red background and light blue background. The trajectories of the electrons recombining at $P_{1}$ are classified as $R_{1}$ and presented by red solid lines. The trajectories of the electrons recombining at $P_{2}$ are classified as $R_{2}$ and presented by blue dotted lines. (b) Laser field vector of the $\omega-2\omega$ field. Bold purple line $d_{i}$ represents the electronic field vector at the ionization time at the first laser lobe. Bold red line at $P_{1}$ and bold blue line at $P_{2}$ represent the electronic laser field vector at the recombination time at the following laser lobes.}
\end{figure}

To give a clear interpretation of the quantum results above, we explore the classical behavior of the electron path by solving Newton's equation under the BCCP laser field driven \cite{Meng,Zhuo,Yin,Tong,Min}. In Fig.2(b), we present the electrons classical ionization times labelled by the blue diamonds, as well as, the recombination times labelled by the red pentagrams. The solid markers are used for the short trajectories and the hollow markers are used for the long trajectories. Note that HHG in bicircular fields is dominated by short trajectories \cite{DB}. We focus on the short trajectories of the electrons ionized at one laser field lobe. As shown in this figure, three ionization bursts and three recombination bursts in one fundamental laser cycle reflect the threefold symmetry of the $\omega-2\omega$ counter-rotating circularly polarized laser field \cite{Long,Eichmann,DB}. The electrons ionized in one laser lobe can recombine with the parent ion in the following two lobes and two peaks marked as $Q_{1}$ and $Q_{2}$ are generated respectively at laser lobe $P_{1}$ and $P_{2}$ (see Fig.3(b)). The maxima of the peak $Q_{1}$ and peak $Q_{2}$ are 111th-order and 122nd-order harmonics respectively, which approximately correspond to the reversion position and the cutoff position. The slight deviation between the quantum and classical model originates from the quantum effect and the absence of harmonic 3n$\omega$ in quantum model. But it does not influence greatly on our conclusion. By comparing the time-frequency analysis and the classical calculations, one can see that, the DCP between the ionization threshold and about 40th-order harmonic is positive, which is only contributed by $Q_{1}$. However, between the maximum of $Q_{1}$ and the harmonic cutoff, the DCP contributed by only $Q_{2}$ is negative. Besides, at the overlapping part of the two peaks $Q_{1}$ and $Q_{2}$ (40th-order harmonic to 111th-order harmonic), the DCP decreases along with the increase of harmonic order. The correspondence relationship between the time-frequency analysis and the classical calculations indicates that the harmonics generated by $Q_{1}$ are left-circularly polarized while the harmonics generated by $Q_{2}$ are right-circularly polarized. Therefore, the harmonics, which are contributed by the electrons ionized at one field lobe and recombining at the following adjacent lobes, have opposite helicities.

To further illustrate our idea, we present the electron classical trajectories and the plot of the laser field vector of the $\omega-2\omega$ field in Fig.3. The dash-dot green line in Fig.3(a) is the amplitude of the laser field with three peaks per optical cycle. Every peak corresponds to one laser lobe. In order to increase the readability of the figure, the second lobe marked as $P_{1}$ and the third lobe marked as $P_{2}$ are superimposed by the light red background and light blue background respectively. We calculate the short trajectories of the electrons ionized at first laser lobe. It is found that, these trajectories can be classified by the laser lobe they recombine at. That is, the trajectories of the electrons recombining at $P_{1}$ are classified as group $R_{1}$ (the red lines), and the trajectories of the electrons recombining at $P_{2}$ are classified as group $R_{2}$ (the blue dot lines). To give a clear description of the trajectory characteristics, the polar plot of the electric field vector of the $\omega$-$2\omega$ field is shown in Fig.3(b). A clover-leaf structure that has a threefold symmetry is presented. The laser field symmetry is reflected by an attosecond pulse train with three pulses generated in one optical cycle in Fig.2. The electronic field vector at the ionization time at the first laser lobe is marked as bold purple line $d_{i}$. The bold red line at $P_{1}$ and bold blue line at $P_{2}$ denote the laser field vector at the recombination times for electrons of groups $R_{1}$ and $R_{2}$ respectively. The electrons ionized at $d_{i}$ can recombine with the nucleus at $P_{1}$ and $P_{2}$ through the path $R_{1}$ and $R_{2}$. Note that the trajectories $R_{1}$ and $R_{2}$ correspondingly contribute to short trajectories of $Q_{1}$ and $Q_{2}$ in Fig.2(b). Then, based on the discussion about Fig.2, the harmonics radiated from $P_{1}$ and $P_{2}$ have opposite helicities.

Next, we also calculate the relative phase of the $x$,$y$ harmonic components using the strong-field approximation (SFA) model \cite{Zhou,Lewenstein}. Fig.4(a) and (b) respectively exhibit the calculation results of the harmonic emissions generated by electrons recombining at $P_{1}$ and $P_{2}$. Based on the classical calculations in Fig.2(b), we choose the recombination time period and integral over a ionized time period for each recombination time chosen. From this figure, one can see that the relative phase of the two components radiated at the laser lobe $P_{1}$ varies between 0.4$\pi$ to $\pi$, which indicates that the radiated harmonics have the positive DCP. However, the relative phase of the two components radiated at the lobe $P_{2}$ changes from 0 to -0.5$\pi$ which indicates that the harmonics have the negative DCP. The result of Fig.4 coincides with that of Fig.2. Therefore, our conclusion that the harmonics contributed by electrons ionized at the first lobe and recombining at the following two lobes have opposite helicities is confirmed by using the classical model and the SFA model.

\begin{figure}
\centerline{
\includegraphics[width=8cm]{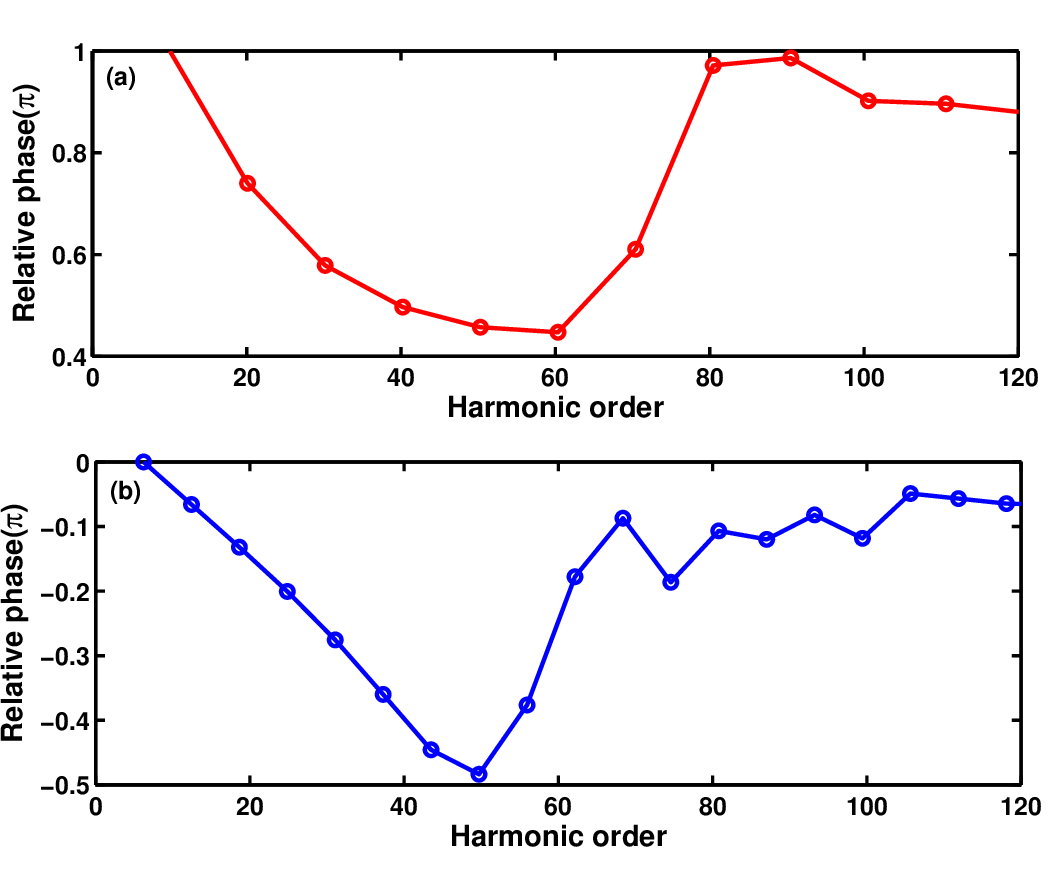}}
\caption{The relative phase of the $x$, $y$ harmonic components radiated from (a) group $R_{1}$ and (b) group $R_{2}$ as the function of harmonic order.}
\end{figure}

\begin{figure}
\centerline{
\includegraphics[width=6cm]{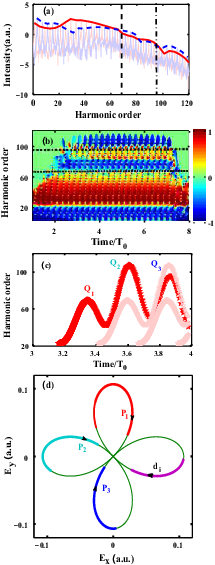}}
\caption{(a) Harmonic intensity as a function of the harmonic order. The spectra for harmonics co-rotating and counter-rotating with the fundamental field are shown as the light red and blue lines respectively. The envelopes of these harmonics are presented as the dark solid red and blue dashed lines correspondingly. (b) The time-frequency distribution of the harmonic DCP. (c) The classical recombination paths in the BCCP laser field. (d) The laser field vector of the $\omega-3\omega$ counter-rotating circularly polarized laser field on the polarization plane. The laser parameters adopted here are $\lambda_{1}=1300$ nm, $\textit{I}_{1}=1.0\times 10^{14}$ W/cm$^2$, $\gamma=3$.}
\end{figure}

Finally, we have also calculated the HHG driven by BCCP laser field with $\gamma=3$. The fundamental laser wavelength and laser intensity are $\lambda_{1}=1300$ nm and $\textit{I}_{1}=1.0\times 10^{14}$ W/cm$^2$, respectively. In Fig.5(a), the harmonic spectra for left-circularly polarized harmonics (4n+1)$\omega$ and right-circularly polarized harmonics (4n-1)$\omega$ are presented with light red and blue lines respectively. The envelopes of these harmonic spectra are presented with the dark red solid lines and blue dashed lines correspondingly. In this spectrum, four helicity reversions are obviously observed. The first one is also at the ionization threshold, which is determined by the target. The remaining three reversions are at about 69th-order (the dashed line), 80th-order and 95th-order (the dash-dot line). In order to clearly display the phenomenon of helicity reversion of the harmonics, we plot the time-frequency distribution of the harmonic DCP. Four attosecond pulses are generated per optical cycle, which reflects the fourfold symmetry of the $\omega-3\omega$ laser field. The helicity reversions at about 69th-order and 95th-order are labelled by the dashed line and the dash-dot line. Similarly, the classical electron paths in one optical cycle are exhibited in Fig.5(c). We only study the trajectories of the electrons ionized at one laser lobe and recombining at the following three lobes. The recombination times of these electrons are labelled by the dark red pentagrams. Others are labelled by the light red pentagrams. Three peaks $Q_{1}, Q_{2}$ and $Q_{3}$ are presented in Fig.5(c). Comparing Fig.5(b) with Fig.5(c), it is shown that, the DCP in the region between the ionization threshold 22nd-order and the reversion position 69th-order is positive and decreases with harmonic order. This region is dominated by $Q_{1}$. The region between the reversion position 95th-order (the maximum of peak $Q_{3}$) and the harmonic cutoff 109th-order (the maximum of peak $Q_{2}$) is only contributed by $Q_{2}$. The helicity of $Q_{2}$ is negative, which is opposite to that of $Q_{1}$. The helicity reversions at the 69th-order and the 95th-order in Figs.5(a) and 5(b) appear at the maxima of $Q_{1}$ and $Q_{3}$ in Fig.5(c). In the overlapping region of $Q_{2}$ and $Q_{3}$ between 69th-order and 95th-order, the harmonic emissions have both positive and negative DCP and an additional helicity reversion appears. This will be discussed later. For a clear analysis of the classical electron trajectories in the $\omega-3\omega$ laser field, the laser field vector is plotted in Fig.5(d). The electronic field vector at the ionization time at the first laser lobe is marked as bold purple line $d_{i}$. The recombination times of the short trajectories of $Q_{1}$, $Q_{2}$ and $Q_{3}$ are marked as the red bold line in laser lobe $P_{1}$, cyan bold line in lobe $P_{2}$ and blue bold line in lobe $P_{3}$ respectively. It is found that the peaks $Q_{1}$, $Q_{2}$ and $Q_{3}$ are contributed by the electrons recombining at different laser lobes. Based on the above comparison between Fig.5(b) and Fig.5(c), the harmonic emissions generated in $P_{1}$ have opposite helicities with those in $P_{2}$, and the helicities of harmonics generated in $P_{2}$ are opposite to those in $P_{3}$, which is consistent with the conclusion discussed above for $\omega-2\omega$ laser field.

To understand the appearance of the additional helicity reversion in the overlapping region of $Q_{2}$ and $Q_{3}$, the weights of the contributing electron trajectories should be obtained. Fig. 6 shows the weights of the short trajectories in both the $\omega-3\omega$ and $\omega-2\omega$ laser fields. For each trajectory, the weight is calculated by $W(t_0, v_\perp)=W_0(t_0)W_1(v_\perp)$. $W_0(t_0)$ is the ionization rate at time $t_0$ obtained from the TDSE simulation by calculating the derivative of the time-dependent ionization probability. $W_1(v_\perp)$ is the weight with a Gaussian distribution for the tunnel electron with transverse initial velocity $v_\perp$ \cite{Delone}. For the $\omega-3\omega$ laser field as shown in Fig. 6(a), the spectrum range can be divided into four regions. In the first region below the 69th-order, the weight of peak $Q_1$ is much greater than those of peaks $Q_2$ and $Q_3$. In the second region from the 69th-order to the 79th-order, peak $Q_1$ disappears and the weight of $Q_2$ is greater than $Q_3$. Then from about 79th-order to 95th-order, the weight of $Q_3$ becomes greater than $Q_2$ in the third region. In the last region above the 95th-order, only peak $Q_2$ remains contributing to the harmonic emissions. In each region, the helicity of the generated harmonics is determined by helicity of the strongest contributing peak, which can well explain the results in Figs. 5(a-b). Owing to the intersection between the weights of peaks $Q_2$ and $Q_3$, the additional helicity reversion in the overlapping region of $Q_{2}$ and $Q_{3}$ appears. For the $\omega-2\omega$ laser field as shown in Fig. 6(b), the spectrum range can be divided by the 111th-order into two regions. The weight of peak $Q_1$ is much greater than that of peak $Q_2$ in the first region. In the second region, only peak $Q_2$ contributes to the harmonic emissions. Since the weights of the two peaks have no intersection, there is no extra helicity reversion in the $\omega-2\omega$ BCCP laser field.

\begin{figure}
\centerline{
\includegraphics[width=8.5cm]{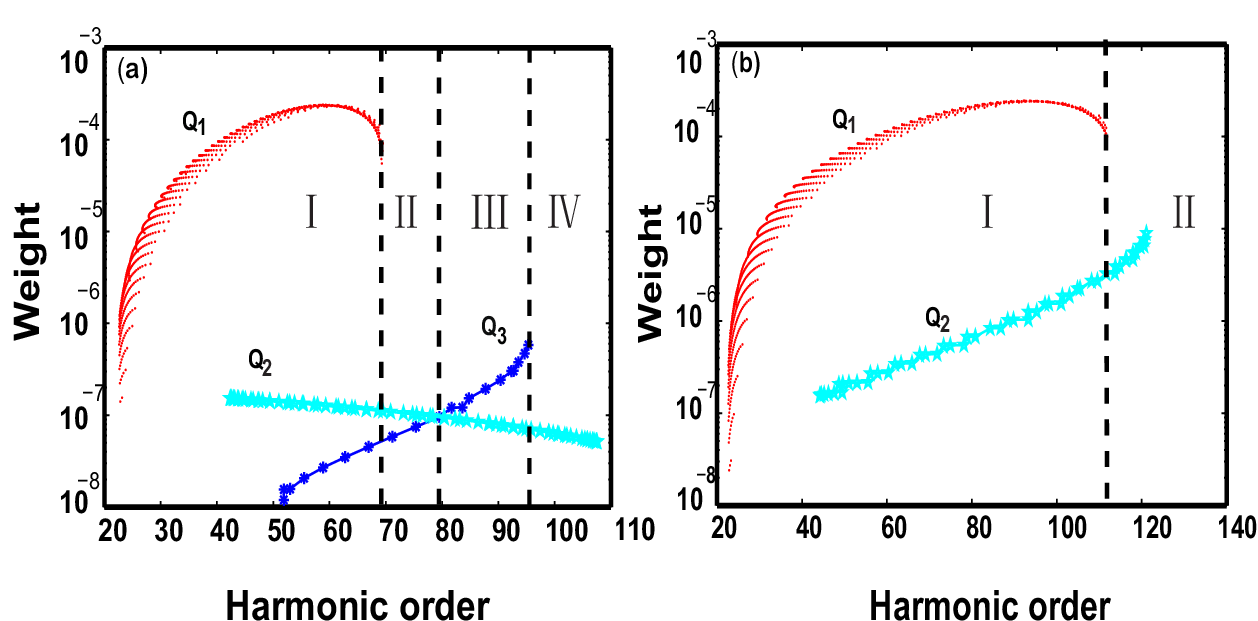}}
\caption{Weights of the contributing short trajectories in the (a) $\omega-3\omega$ and (b) $\omega-2\omega$  BCCP laser fields. The red dots, cyan pentagrams present the weights for peaks $Q_1$, $Q_2$ for $\omega-3\omega$ and $\omega-2\omega$ laser fields, and the blue stars presents the weights for peak $Q_3$ for $\omega-3\omega$ laser field.}
\end{figure}

Based on all the discussions above, a picture of the electron classical trajectories in HHG driven by BCCP laser field is revealed. Besides, the correspondence between the harmonic polarization and the classical trajectories has also been established. Specifically, the electron ionized at one laser lobe can recombine with the nucleus at the following laser lobes. These recombining trajectories can be classified into different groups based on the laser lobe they recombining at. Moreover, the harmonics emitted from these groups at adjacent laser lobes have opposite helicities. For instance, in $\omega-2\omega$ laser field, groups $R_{1}$ and $R_{2}$ are both ionized at one laser lobe $d_{i}$. And they recombine with the nucleus at the following laser lobes $P_{1}$ and $P_{2}$ and form the peaks $Q_1$ and $Q_2$, respectively. The harmonics from $Q_{1}$ and $Q_{2}$ are radiated with opposite helicities and the helicity reversion occurs at the maximum of $Q_1$. In $\omega-3\omega$ laser field, electrons ionized at one laser lobe $d_{i}$ recombine at the following lobes $P_{1}$, $P_{2}$, $P_{3}$ and form the peaks $Q_1$, $Q_2$, $Q_3$, respectively. The harmonics from peaks $Q_1$ and $Q_3$ have positive helicity while those from $Q_2$ has negative helicity. The helicity reversion occurs at the position where the dominantly contributing peak is changed to its neighbouring peak, i.e. at the maxima of $Q_1$ and $Q_2$ and at the intersection between the weights of $Q_2$ and $Q_3$.

\section{Conclusion}
In conclusion, under the interaction of bicircular laser field with the 2D model Ne atoms, a harmonic spectrum with different helicities in different spectrum ranges is generated. It is shown that, the helicity of the elliptically polarized harmonic emission is reversed at particular harmonic orders. Based on the time-frequency analysis and the classical three-step model, the relationship between the reversion position and the electron trajectories  is obtained. The electrons ionized at one lobe of laser field can recombine with the nucleus at the following laser lobes radiating harmonics. These harmonic emissions generated from adjacent laser lobes have opposite helicities. Besides, we confirm our conclusion through the calculation of relative phase by the SFA model. Our study performs a detailed analysis for the HHG driven by the BCCP laser field in terms of the classical trajectories and established the correspondence between the harmonic polarization and the classical trajectories.

\section*{Acknowledgement}

This work was supported by the National Natural Science Foundation of China under Grants No. 11404123, 11234004, 61275126, 11422435 and 11574101. Numerical simulations presented in this paper were carried out using the High Performance Computing Center experimental testbed in SCTS/CGCL (see http://grid.hust.edu.cn/hpcc).

\end{document}